\documentclass[sigconf,screen]{acmart}

\AtBeginDocument{%
  \providecommand\BibTeX{{%
    \normalfont B\kern-0.5em{\scshape i\kern-0.25em b}\kern-0.8em\TeX}}}

\setcopyright{rightsretained}
\acmPrice{}
\acmDOI{10.1145/3611643.3616257}
\acmYear{2023}
\copyrightyear{2023}
\acmSubmissionID{fse23main-p132-p}
\acmISBN{979-8-4007-0327-0/23/12}
\acmConference[ESEC/FSE '23]{Proceedings of the 31st ACM Joint European Software Engineering Conference and Symposium on the Foundations of Software Engineering}{December 3--9, 2023}{San Francisco, CA, USA}
\acmBooktitle{Proceedings of the 31st ACM Joint European Software Engineering Conference and Symposium on the Foundations of Software Engineering (ESEC/FSE '23), December 3--9, 2023, San Francisco, CA, USA}

\usepackage[bottom]{footmisc}
\usepackage{multirow}
\usepackage{booktabs} 
\usepackage{textcomp}
\usepackage{xcolor}
\usepackage{graphicx}
\usepackage{caption}
\usepackage{graphics}
\usepackage{rotating}
\usepackage{tabularx}
\usepackage{wrapfig}
\usepackage{listings}
\usepackage{color, colortbl}
\usepackage{balance} 
\usepackage{lscape}
\usepackage{hyperref}
\usepackage{xspace}
\usepackage{framed}
\usepackage{syntax}
\usepackage{soul}
\usepackage{flushend}
\usepackage{caption}
\usepackage{subcaption}
\usepackage{tabularx,ragged2e}
\usepackage{booktabs}
\usepackage{amsmath}
\usepackage{algorithm}
\usepackage[noend]{algpseudocode}
\usepackage{url}
\usepackage{tabularx}
\usepackage{array}
\usepackage{colortbl}
\usepackage{varwidth}
\usepackage[justification=centering]{caption}
\usepackage[tikz]{bclogo}
\usepackage[braket]{qcircuit}
\usepackage{amsmath}  
\usepackage{microtype}

\definecolor{cellgreen}{HTML}{D8FFD7}
\definecolor{cellblue}{HTML}{ECF4FF}
\definecolor{cellyellow}{HTML}{FFFFC7}
\definecolor{cellgrey}{HTML}{EFEFEF}
\definecolor{cellpink}{HTML}{FFE2E1}

\makeatletter
\def\BState{\State\hskip-\ALG@thistlm}
\makeatother
\definecolor{sandstorm}{rgb}{0.93, 0.84, 0.25}
\definecolor{codegreen}{rgb}{0,0.6,0}
\definecolor{codegray}{rgb}{0.5,0.5,0.5}
\definecolor{codepurple}{rgb}{0.58,0,0.82}
\definecolor{backcolour}{rgb}{0.95,0.95,0.92}
\definecolor{Gray}{gray}{0.1}
\definecolor{Blanched Almond}{rgb}{1.0, 0.92, 0.8}
\definecolor{darksalmon}{rgb}{0.91, 0.59, 0.48}
\definecolor{AliceBlue}{rgb}{0.94, 0.97, 1.0}
\definecolor{darkred}{rgb}{0.55, 0.0, 0.0}
\definecolor{Steelblue}{rgb}{0.27, 0.51, 0.71}

\newcolumntype{Y}{>{\raggedleft\arraybackslash}X}

\lstdefinestyle{mystyle}{
	backgroundcolor=\color{AliceBlue},   
	commentstyle=\color{codegreen},
	keywordstyle=\color{magenta},
	numberstyle=\tiny\color{codegray},
	stringstyle=\color{codepurple},
	basicstyle=\footnotesize,
	breakatwhitespace=false,         
	breaklines=true,                 
	captionpos=b,                    
	keepspaces=true,                 
	numbers=left,                    
	numbersep=5pt,                  
	showspaces=false,                
	showstringspaces=false,
	showtabs=false,                  
	tabsize=2,
	columns=fullflexible,
	moredelim=[is][\color{blue}\bfseries\underbar]{@}{@},
	escapechar=\@,
	mathescape=true
}

\newcommand{\as}{\textit{Auto-Sklearn}\xspace}

\newcommand{\kg}{\textit{Kaggle}\xspace}

\newcommand{\etal}{{\em et al.}\xspace}
\newcommand{\fairautoml}{\textit{Fair-AutoML}\xspace}
\newcommand{\fairea}{\textit{Fairea}\xspace}
\newcommand{\openml}{\textit{OpenML}\xspace}

\newcounter{NumObservations}
\stepcounter{NumObservations}

\begin{document}

\title{Fix Fairness, Don't Ruin Accuracy: Performance Aware Fairness Repair using AutoML}

\author{Giang Nguyen}
\email{gnguyen@iastate.edu}
\affiliation{%
  \institution{Dept. of Computer Science\\Iowa State University}
  \city{Ames}
  \state{Iowa}
  \country{USA}
}

\author{Sumon Biswas}
\email{sumonb@cs.cmu.edu}
\affiliation{
	\institution{School of Computer Science\\Carnegie Mellon University}
	\city{Pittsburgh}
	\state{PA}
	\country{USA}
	}

\author{Hridesh Rajan}
\email{hridesh@iastate.edu}
\affiliation{%
  \institution{Dept. of Computer Science\\Iowa State University}
  \city{Ames}
  \state{Iowa}
  \country{USA}
}


\begin{abstract}

Machine learning (ML) is increasingly being used in critical decision-making software, but incidents have raised questions about the fairness of ML predictions. To address this issue, new tools and methods are needed to mitigate bias in ML-based software. Previous studies have proposed bias mitigation algorithms that only work in specific situations and often result in a loss of accuracy. Our proposed solution is a novel approach that utilizes automated machine learning (AutoML) techniques to mitigate bias. Our approach includes two key innovations: a novel optimization function and a fairness-aware search space. By improving the default optimization function of AutoML and incorporating fairness objectives, we are able to mitigate bias with little to no loss of accuracy. Additionally, we propose a fairness-aware search space pruning method for AutoML to reduce computational cost and repair time. Our approach, built on the state-of-the-art \as tool, is designed to reduce bias in real-world scenarios. In order to demonstrate the effectiveness of our approach, we evaluated our approach on four fairness problems and 16 different ML models, and our results show a significant improvement over the baseline and existing bias mitigation techniques. Our approach, \fairautoml, successfully repaired 60 out of 64 buggy cases, while existing bias mitigation techniques only repaired up to 44 out of 64 cases.

\end{abstract}

\ccsdesc[500]{Software and its engineering~Search-based software engineering}
\ccsdesc[500]{Computing methodologies~Machine learning}

\keywords{Software fairness, bias mitigation, fairness-accuracy trade-off, machine learning software, automated machine learning
}



\maketitle

\section{Introduction}
\label{sec:intro}

Recent advancements in machine learning have led to remarkable success in solving complex decision-making problems such as job recommendations, hiring employees, social services, and education~\cite{byanjankar2015predicting, bogen2018help, hoffman2018discretion, malekipirbazari2015risk, he2014practical, perlich2014machine, chouldechova2018case, vaithianathan2013children, de2018clinically, kourou2015machine, eubanks2018automating, oneto2019learning, ahmed23dlcontract}. However, ML software can exhibit discrimination due to unfairness bugs in the models~\cite{angwin2016machine, biswas20machine}. These bugs can result in skewed decisions towards certain groups of people based on protected attributes such as race, age, or sex~\cite{friedler2019comparative, galhotra2017fairness}. 

To address this issue, the software engineering (SE) community has invested in developing testing and verification strategies to detect unfairness in software systems~\cite{biswas23fairify, friedler2019comparative, aggarwal2019black, galhotra2017fairness, udeshi2018automated}. Additionally, the machine learning literature contains a wealth of research on defining different fairness criteria for ML models and mitigating bias~\cite{calders2010three, chouldechova2017fair, feldman2015certifying, hardt2016equality, kamiran2012data, kamishima2012fairness,  zafar2017fairness, zhang2018mitigating}. Various bias mitigation methods have been proposed to build fairer models. Some approaches mitigate data bias by adapting the training data~\cite{li2022training, chakraborty2020fairway, chakraborty2021bias}; some modify ML models during the training process to mitigate bias~\cite{tao2022ruler, chen2022maat, hort2021did, gao2022fairneuron, tizpaz2022fairness}, and others aim to increase fairness by changing the outcome of predictions~\cite{aggarwal2019black, udeshi2018automated, zhang2020white}.

Despite these efforts, current bias mitigation techniques often come at the cost of decreased accuracy~\cite{hort2021fairea, biswas20machine}. Their effectiveness varies based on datasets, fairness metrics, or the choice of protected attributes~\cite{feffer2022empirical, feldman2015certifying, hardt2016equality, chen2022comprehensive}. \citeauthor{hort2021fairea} proposed Fairea~\cite{hort2021fairea}, a novel approach to evaluate the effectiveness of bias mitigation techniques, which found that nearly half of the evaluated cases received poor effectiveness. Moreover, evaluations by Chen \etal also showed that in 25\% of cases, bias mitigation methods reduced both ML performance and fairness~\cite{chen2022comprehensive}. 

Recent works~\cite{hort2021fairea, tizpaz2022fairness, gohar23fairness} have shown that parameter tuning can successfully fix fairness bugs without sacrificing accuracy. By finding the best set of parameters, parameter tuning can minimize the error between the predicted values and the true values to reduce bias. This helps to ensure that the model is not overly simplified or too complex, which can lead to underfitting (high bias) or overfitting (low accuracy), respectively. By tuning the parameters, we can find the right balance between bias and accuracy, which leads to a model that generalizes well to different data or fairness metric. However, it is challenging to identify which parameter setting achieves the best fairness-accuracy trade-off~\cite{gohar23fairness}. 

Recent advancements in AutoML technology~\cite{feurer2015efficient, feurer2020auto, jin2019auto} have made it possible for both experts and non-experts to harness the power of machine learning. AutoML proves to be an effective option for discovering optimal parameter settings; however, currently there is a lack of focus on reducing bias within the AutoML techniques. Thus, we pose the following research questions: \emph{Is it possible to utilize AutoML for the purpose of reducing bias? Is AutoML effective in mitigating bias? Does AutoML outperform existing bias reduction methods? Is AutoML more adaptable than existing bias mitigation techniques?}

We introduce \fairautoml, a novel technique that utilizes AutoML to fix fairness bugs in machine learning models. Unlike existing bias mitigation techniques, \fairautoml addresses their limitations by enabling efficient and fairness-aware Bayesian search to repair unfair models, making it effective for a wide range of datasets, models, and fairness metrics. The key idea behind \fairautoml is to use AutoML to explore as many configurations as possible in order to find the optimal fix for a buggy model. Particularly, \fairautoml enhances the potential of AutoML for fixing fairness bugs in two novel techniques: by generating a new optimization function that guides AutoML to fix fairness bugs without sacrificing accuracy, and by defining a new search space based on the specific input to accelerate the bug-fixing process. Together, these contributions enable \fairautoml to effectively fix fairness bugs across various datasets and fairness metrics. We have implemented \fairautoml on top of \as~\cite{feurer2015efficient}, the state-of-the-art AutoML framework. 




\fairautoml aims to effectively address the limitations of existing bias mitigation techniques by utilizing AutoML to efficiently repair unfair models across various datasets, models, and fairness metrics. We conduct an extensive evaluation of \fairautoml using 4 widely used datasets in the fairness literature~\cite{galhotra2017fairness, udeshi2018automated, aggarwal2019black} and 16 buggy models collected from a recent study~\cite{biswas20machine}. The results demonstrate the effectiveness of our approach, as \fairautoml successfully repairs 60 out of 64 buggy cases, surpassing the performance of existing bias mitigation techniques which were only able to fix up to 44 out of 64 bugs in the same settings and training time.


Our main contributions are the following:
\begin{itemize}
		\item We have proposed a novel approach to fix unfairness bugs and retain accuracy at the same time.
		\item We have proposed methods to generate the optimization function automatically based on an input to make AutoML fixing fairness bugs more efficiently.
		\item We have pruned the search space automatically based on an input to fix fairness bugs faster using AutoML.
		\item We have implemented our approach in a SOTA AutoML, 
		\as~\cite{feurer2015efficient}. 
		{\bf The artifact is available here~\cite{fazenodo}}.
\end{itemize}

The paper is organized as follows: \S\ref{sec:background} describes the background, \S\ref{sec:motiv} presents a motivation, \S\ref{sec:problem} indicates the problem definition, \S\ref{sec:method} shows the \fairautoml approaches, \S\ref{sec:eval} presents the our evaluation, \S\ref{sec:discussion} discusses the limitations and future directions of \fairautoml, \S\ref{sec:limit} discusses the threats to validity of \fairautoml, \S\ref{sec:conclusion} concludes, and \S\ref{sec:data} describes the artifact.

\section{Background}
\label{sec:background}
We begin by providing an overview of the background and related research in the field of software fairness.
\subsection{Preliminaries}
\subsubsection{ML Software} Given an input dataset $D$ split into a training dataset $D_{train}$ and a validation dataset $D_{val}$, a ML software system can be abstractly viewing  as mapping problem $M_{\lambda,c}$ : $x$ $\rightarrow$ $y$ from inputs $x$ to outputs $y$ by learning from $D_{train}$. ML developers aims to search for a hyperparameter configuration $\lambda*$ and complementary components $c*$ for model $M$ to obtain optimal fairness-accuracy on $D_{val}$. The complementary components can be ML algorithms combined with a classifier i.e., pre-processing algorithms. 

\subsubsection{AutoML} 
Given the search spaces $\Lambda$ and $C$ for hyperparameters and complementary components, AutoML aims to find $\lambda*$ and $c*$ to obtain the lowest value of the cost function (Equation \ref{eq:cost1}):
\begin{equation}
    M = \underset{\mathcal{\lambda} \in \Lambda, c \in C}{\arg \min}Cost(M_{\lambda*,c*}, D_{val})
	\label{eq:cost1}
\end{equation}
\begin{equation}
    (\lambda*,c*) = \underset{(\lambda, c)}{\arg \min}Loss(M_{\lambda,c}, D_{train})
	\label{eq:loss}
\end{equation}

\subsubsection{Measures} We consider a problem, where each individual in the population has a true label in $y$ = \{0, 1\}. We assume a protected attribute $z$ = \{0, 1\}, such as race, sex, age, where one label is privileged (denoted 0) and the other is unprivileged (denoted 1). The predictions are $\hat{y} \in \{0, 1\}$ that need to be not only accurate with respect to $y$ but also fair with respect to the protected attribute $z$. 
\paragraph{Accuracy Measure} Accuracy is given by the ratio of the number of correct predictions by the total number of predictions. 

\begin{center}
	Accuracy = (\# True positive + \# True negative) / \# Total
\end{center}	

\paragraph{Fairness Measure} We use four ways to define group fairness metrics, which are widely used in fairness literature~\cite{bellamy2018ai, friedler2019comparative, binns2018fairness}:
\subparagraph{The \emph{Disparate Impact (DI)} is the proportion of the unprivileged group with the favorable label divided by the proportion of the privileged group with the favorable label~\cite{feldman2015certifying, zafar2017fairness}.}
\begin{center}
	$DI = \frac{Pr[\hat{y}=1|z=0]}{Pr[\hat{y}=1|z=1]}$
\end{center}

\subparagraph{The \emph{Statistical Parity Difference (SPD)} quantifies the disparity between the favorable label's probability for the unprivileged group and the favorable label's probability for the privileged group~\cite{calders2010three}.}
\begin{center}
\scalebox{0.9}	{$SPD = Pr[\hat{y}=1|z=0] - Pr[\hat{y}=1|z=1]$}
\end{center}	

\subparagraph{The \emph{Equal Opportunity Difference} (EOD) measures the disparity between the true-positive rate of the unprivileged group and the privileged group.}
\begin{center}
\scalebox{0.9} {$TPR_u = Pr[\hat{y} = 1|y = 1,z = 0];  TPR_p = Pr[\hat{y} = 1|y = 1,z = 1]$}
 
\scalebox{0.9} {$EOD = TPR_u - TPR_p$}
\end{center}	
\subparagraph{
\emph{The Average Absolute Odds Difference (AOD)} is the mean of the difference of true-positive rate and false-positive rate among the unprivileged group and privileged group~\cite{hardt2016equality}.}
\begin{center}
\scalebox{0.9} {$FPR_u = P[\hat{y} = 1|y = 0,z = 0];  FPR_p = P[\hat{y} = 1|y = 0,z = 1]$}
 
\scalebox{0.9}{$AOD = \frac{1}{2} * {|FPRu - FPRp| + |TPR_u - TPR_p|}$}
\end{center}

\subparagraph{To use all the metrics in the same setting, DI has been
plotted in the absolute value of the log scale, and SPD, EOD, AOD have been plotted in absolute value~\cite{chakraborty2020fairway, hort2021fairea}. Thus, the bias score of a model is measured from 0, with lower scores indicating more fairness.}

\subsection{Related Work} 
\subsubsection{Bias mitigation}
SE and ML researchers has developed various bias mitigation methods to increase fairness in ML software divided into three categories~\cite{hort2022bia, friedler2019comparative}:

\textbf{Pre-processing} approaches reduce bias by pre-processing the training data. For instance, \emph{Fair-SMOTE}~\cite{chakraborty2021bias} addresses data bias by removing biased labels and balancing the distribution of positive and negative examples for each sensitive attribute. \emph{Reweighing}~\cite{kamiran2012data} decreases bias by assigning different weights to different groups based on the degree of favoritism of a group. \emph{Disparate Impact Remover}~\cite{feldman2015certifying} is a pre-processing bias mitigation technique that aims to reduce bias by editing feature values. 

\textbf{In-processing} approaches reduce bias by modifying ML models during the training process i.e., \emph{Parfait-ML}~\cite{tizpaz2022fairness} present a search-based solution to balance fairness and accuracy by tuning hyperparameter to approximate the twined Pareto curves. \emph{MAAT}~\cite{chen2022maat} is an ensemble approach aimed at improving the fairness-performance trade-off in ML software. Instead of combining models with the same learning objectives as traditional ensemble methods, \emph{MAAT} merges models that are optimized for different goals.

\textbf{Post-processing} approaches change the outcome of prediction to reduce bias. This technique unfavors privileged groups' instances and favors those of unprivileged groups lying around the decision boundary. For example, \emph{Equalized Odds}~\cite{hardt2016equality} reduces the value of EOD by modifying the output labels. \emph{Fax-AI}~\cite{grabowicz2022marrying} eliminates direct discrimination in machine learning models by limiting the use of certain features, thereby preventing them from serving as surrogates for protected attributes.
 \emph{Reject Option Classification}~\cite{kamiran2012decision} prioritizes instances from the privileged group over those from the unprivileged group that are situated on the decision boundary with high uncertainty.

Previous efforts have made significant progress in reducing bias; however, they come at the cost of decreased accuracy and their results can vary depending on the datasets and fairness metrics. Our proposal, \fairautoml, aims to strike a balance between accuracy and bias reduction and demonstrate generalizability across various datasets and metrics.


\subsubsection{Search space pruning} 
Search space pruning involves reducing the size or complexity of the search space in optimization or machine learning tasks. Pruning techniques are employed to accelerate the optimization process of AutoML by eliminating unpromising or redundant options, thus focusing computational resources on more promising areas of the search space. For example, Feurer \etal~\cite{feurer2015efficient} introduce Auto-Sklearn 2.0, a novel approach aimed at enhancing the performance of Auto-Sklearn. This advancement involves constraining the search space to exclusively comprise iterative algorithms, while eliminating feature preprocessing. This strategic adjustment streamlines the implementation of successive halving, as it reduces the complexity to a single fidelity type: the number of iterations. Otherwise, the incorporation of dataset subsets as an alternative fidelity would require additional consideration. Another innovative contribution comes from Cambronero \etal, who introduces AMS~\cite{cambronero2020ams}. This method capitalizes on the wealth of source code repositories to streamline the search space for AutoML. Notably, AMS harnesses the power of unspecified complementary and functionally related API components. By leveraging these components, the search space for AutoML is pruned effectively. Diverging from prior research efforts, \fairautoml distinguishes itself by leveraging data characteristics to effectively trim down the search space. Notably, existing techniques in search space pruning primarily target accuracy enhancement within AutoML. In contrast, our innovative pruning methodology within \fairautoml is uniquely directed towards repairing unfair models.

\subsubsection{AutoML extension} 
AutoML aims to automate the process of building a high-performing ML model, but it has limitations. It can be costly, time-consuming to train, and produces complex models that are difficult to understand. To address these limitations, software engineering researchers have developed methods to enhance AutoML performance, such as \emph{AMS}~\cite{cambronero2020ams} and \emph{Manas}~\cite{nguyen2021manas}. \emph{AMS} utilizes source code repositories to create a new search space for AutoML, while \emph{Manas} mines hand-developed models to find a better starting point for AutoML. The goal of these methods is to improve AutoML to maximize the accuracy. Different from these methods, \fairautoml, built on top of \as~\cite{feurer2015efficient}, is the first to focus on repairing unfair models.

\section{Motivation}
\label{sec:motiv}

The widespread use of machine learning in software development has brought attention to the issue of fairness in ML models. Although various bias mitigation techniques have been developed to address this issue, they have limitations. These techniques suffer from a poor balance between fairness and accuracy~\cite{hort2021fairea}, and are not applicable to a wide range of datasets, metrics, and models~\cite{feffer2022empirical, feldman2015certifying, hardt2016equality}. To gain a deeper understanding of these limitations, we evaluate six different bias mitigation techniques using four fairness metrics, four datasets, and six model types. The evaluation criteria are borrowed from Fairea~\cite{hort2021fairea} and are presented in Table \ref{tbl:motivation}. 

 \begin{figure}
    \centering
	\includegraphics[width=0.4\textwidth]{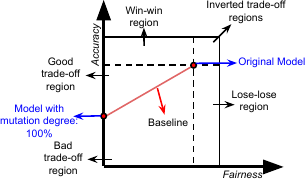}
 \vspace{-1em}
	\caption{Baseline fairness-accuracy trade-off~\cite{hort2021fairea}}
	   \vspace{-1em}
	\label{fig:baseline}
\end{figure}

\fairea is designed to assess the trade-off between fairness and accuracy of bias mitigation techniques. The methodology of \fairea is demonstrated in Figure \ref{fig:baseline}, where the fairness and accuracy of a bias mitigation technique on a dataset are displayed in a two-dimensional coordinate system. The baseline is established by connecting the fairness-accuracy points of the original model and the mitigation models on the dataset. \fairea evaluates the performance of the mitigation technique by altering the original model predictions and replacing a random subset of the predictions with other labels. The mutation degree ranges from 10\% to 100\% with a step-size of 10\%. The baseline classifies the fairness-accuracy trade-off of a bias mitigation technique into five regions: lose-lose trade-off (\textbf{lose}), bad trade-off (\textbf{bad}), inverted trade-off (\textbf{inv}), good trade-off (\textbf{good}), and win-win trade-off (\textbf{win}). A technique reducing both accuracy and fairness would fall into the lose-lose trade-off region. If the trade-off is worse than the baseline, it would fall into the bad trade-off region. If the trade-off is better than the baseline, it would fall into the good trade-off region. If a bias mitigation method simultaneously decreases both bias and accuracy, it would fall into the inverted trade-off region. If the technique improves both accuracy and fairness, it would fall into the win-win trade-off region.

The results of the region classification of six bias mitigation techniques - \emph{Reweighing}~\cite{kamiran2012data}, \emph{Disparate Impact Remover}~\cite{feldman2015certifying}, \emph{Parfait-ML}~\cite{tizpaz2022fairness}, \emph{Equalized Odds}~\cite{hardt2016equality}, \emph{FaX-AI}~\cite{grabowicz2022marrying}, \emph{Reject Option Classification}~\cite{kamiran2012decision} - are shown in Table \ref{tbl:motivation}. The evaluation was conducted on 64 buggy cases using different criteria such as fairness metrics and datasets. The case is identified as buggy when it falls below the \emph{Fairea} baseline. The mean percentage of each technique falling into the corresponding regions is listed in each cell. The mean results provide a general overview of the current state of bias mitigation techniques. Further details on the performance of each individual bias mitigation technique can be found in Table \ref{tbl:percentage} of our evaluation.


Table \ref{tbl:motivation} illustrates that the majority of existing bias mitigation techniques have a poor fairness-accuracy trade-off across different datasets, fairness metrics, and classification models. Specifically, 39\% of the cases show that these techniques perform worse than the original model, with 28\% of the cases resulting in a poor trade-off and 11\% resulting in a decrease in accuracy and an increase in bias. Additionally, Table \ref{tbl:motivation} shows that the performance of these techniques varies depending on the input, as demonstrated by the different results obtained when using different datasets or fainess metrics~\cite{feffer2022empirical, feldman2015certifying, hardt2016equality}. For example, the bias mitigation techniques had a high performance in 62\% of the cases using the Adult dataset (55\% for good trade-off region and 7\% for win-win trade-off region), but only achieved 40\% good effectiveness in the Bank dataset.

\begin{table} 
\centering
\caption{Mean proportions of mitigation cases that
 that fall into each mitigation region}
\setlength{\tabcolsep}{5.8pt}
\vspace{-1em}
\begin{tabular}{|cl|l|l|l|l|l|}
\hline
\multicolumn{2}{|c|}{Criteria}                           & Lose          & Bad           & Inv          & Good          & Win           \\ \hline
\multicolumn{1}{|c|}{\multirow{4}{*}{Metric}}  & DI      & 7\%           & 31\%          & 5\%          & 43\%          & 14\%          \\ \cline{2-7} 
\multicolumn{1}{|c|}{}                         & SPD     & 4\%           & 36\%          & 6\%          & 40\%          & 14\%          \\ \cline{2-7} 
\multicolumn{1}{|c|}{}                         & EOD     & 23\%          & 15\%          & 14\%         & 40\%          & 8\%           \\ \cline{2-7} 
\multicolumn{1}{|c|}{}                         & AOD     & 9\%           & 30\%          & 5\%          & 40\%          & 16\%          \\ \hline
\multicolumn{1}{|c|}{\multirow{4}{*}{Dataset}} & Adult~\cite{adult2017kaggle}   & 18\%          & 6\%           & 14\%         & 55\%          & 7\%           \\ \cline{2-7} 
\multicolumn{1}{|c|}{}                         & Bank~\cite{bank2017kaggle}     & 9\%           & 44\%          & 7\%          & 23\%          & 17\%          \\ \cline{2-7} 
\multicolumn{1}{|c|}{}                         & German~\cite{german2017kaggle}   & 6\%           & 36\%          & 2\%          & 46\%          & 10\%          \\ \cline{2-7} 
\multicolumn{1}{|c|}{}                         & Titanic~\cite{titanic2017kaggle}  & 11\%          & 26\%          & 3\%          & 43\%          & 17\%          \\ \hline
\multicolumn{2}{|c|}{\textbf{Mean}}                      & \textbf{11\%} & \textbf{28\%} & \textbf{7\%} & \textbf{41\%} & \textbf{13\%} \\ \hline
\end{tabular}
\label{tbl:motivation}
\centering
\footnotesize \* Bad: bad trade-off region, Lose: lose-lose trade-off region, Inv: inverted trade-off region, Good: good trade-off region, Win: win-win trade-off region. 
\vspace{-2em}
\end{table}


Hort \etal~\cite{hort2021fairea} have demonstrated that through proper parameter tuning, it is possible to address fairness issues in machine learning models without sacrificing accuracy. However, determining the optimal fairness-accuracy trade-off can be a challenge. Although AutoML can be effective in finding the best parameter settings, it does not specifically address bias reduction. This motivates the development of \fairautoml, a novel approach that utilizes Bayesian optimization to tune parameters and address fairness issues without hindering accuracy. \fairautoml is evaluated for its generality across different fairness metrics and datasets, and unlike other bias mitigation methods, it can be applied to any dataset or metric.


This work focuses on improving fairness quantitatively of buggy models instead of targeting a specific type of datasets and models. Our method is general since we utilize the power of AutoML to try as many configurations as possible to obtain the optimal fix; therefore, our method can work on various types of datasets and metrics. The rest of this work describes our approach, \fairautoml, that addresses the limitations of both existing bias mitigation methods and AutoML. As a demonstration, \fairautoml achieved good performance in 100\% of the 16 buggy cases in the Adult dataset, while 75\% of the mitigation cases showed a good fairness-accuracy trade-off, and the remaining 25\% exhibited an improvement in accuracy without sacrificing bias reduction.

\section{Problem Definition}
\label{sec:problem}


This work aims to utilize AutoML to address issues of unfairness in ML software by finding a new set of configurations for the model that achieves optimal fairness-accuracy trade-off. Because fairness is an additional consideration beyond accuracy, the problem becomes a multi-objective optimization problem, requiring a new cost function that can optimize both fairness and accuracy simultaneously. To achieve this, we use a technique called weighted-sum scalarization (Equation \ref{eq:weightedsum})~\cite{ehrgott2005multicriteria}, which allows us to weigh the importance of different objectives and create a single scalar cost function. 
\vspace{-0.3em}
\begin{equation}
    \vspace{-0.3em}
	A =  \sum_{i=1}^{n} c_{i}*\beta_i  \label{eq:weightedsum}
\end{equation}	
where, $\beta_i$ denotes the relative weight of importance of $c_{i}$:
\vspace{-0.3em}
\begin{equation}
    \vspace{-0.3em}
	\sum_{i=1}^{n} \beta_i = 1
\end{equation}
In this work, we use a cost function (or objective function) that is a weighted-sum scalarization of two decision criteria: bias and accuracy. This cost function, as shown in Equation \ref{eq:optimize}, assign weights to bias and accuracy in the cost function allow us to adjust the trade-off between the two criteria according to the specific problems:
\begin{equation}
Cost(M_{\lambda,c}, \mathcal{D}(z)) = \beta * f + (1 - \beta) * (1 -  a)
\label{eq:optimize}
\end{equation}
We analyze the output of the buggy ML software (including bias and accuracy) to create a suitable cost function for each input. By analyzing the output, we are able to automatically estimate the weights of the cost function in order to balance fairness and accuracy for a specific problem. To the best of our knowledge, this is the first work that applies output analysis of the software to AutoML to repair unfair ML models.

However, using AutoML can be costly and time-consuming. To address this issue, we propose a novel method that automatically create new search spaces $\Lambda*$ and $C*$ based on different inputs to accelerate the bug-fixing process of AutoML. These new search spaces are smaller in size compared to the original ones, $|\Lambda*| < |\Lambda|$ and $|C*| < |C|$. Particularly, as shown in Equation \ref{eq:cost}, \fairautoml takes as input a ML model and a dataset with a protected attribute $z$, and aims to find $\lambda*$ and $c*$ in the smaller search space, in order to minimize the cost value.
\begin{equation}
    M = \underset{\mathcal{\lambda*} \in \Lambda*, c* \in C*}{\arg \min}Cost(M_{\lambda*,c*}, D_{val}(z))
	\label{eq:cost}
\end{equation}
The technique of search space pruning in \fairautoml utilizes data characteristics to enhance bug-fixing efficiency. By shrinking the search spaces based on input analysis, \fairautoml can find better solutions more quickly. A set of predefined modifications to the ML model are pre-built and used as a new search space for new input datasets, reducing the time needed to fix buggy models. Our approach is based on previous works in AutoML~\cite{feurer2015efficient}, but updated and modified to tackle bias issues. To the best of our knowledge, we are the first to propose a search space pruning technique for fairness-aware AutoML.

\section{Fair-AutoML}
\label{sec:method}

\begin{figure}
    \centering
	\includegraphics[width=0.47\textwidth]{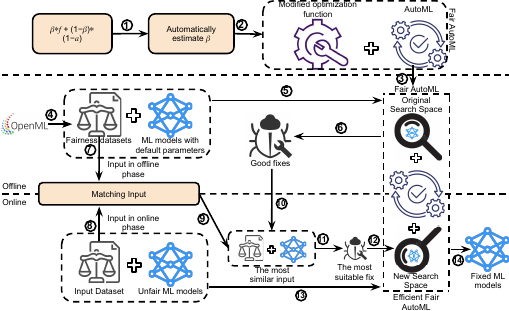}
    \vspace{-0.7em}
    \caption{An Overview of \fairautoml Approach}
    \label{fig:overview}
    \vspace{-0.7em}
\end{figure}

This section describes a detailed description of key components of \fairautoml (Figure \ref{fig:overview}): the dynamic optimization function (steps 1-3) and the search space pruning (steps 4-13).


\subsection{Dynamic Optimization for Bias Elimination}
We strive to eliminate bias in unfair models by utilizing Equation \ref{eq:optimize} as the objective function and determining the optimal value of $\beta$ to minimize the cost function. In this section, we propose an approach to automatically estimate the optimal value of $\beta$ for a specific dataset and a targeted model. This method ensures efficient correction of fairness issues while maintaining high predictive accuracy.



\subsubsection{Upper bound of the cost function} To estimate the optimal value of $\beta$, the first step is to determine the upper bound of the cost function. This can be done by using a "pseudo-model", which is the 100\% mutation degree model~\cite{hort2021fairea}, as shown in the Figure \ref{fig:baseline}. In other words, the pseudo-model always achieves the accuracy on any binary classification problem as follows: 
\begin{equation}
    a_0 = max(P(Y = 1), P(Y = 0))
\end{equation}

Given an input, the pseudo-model achieves an accuracy of $a_0$ and a bias value of $f_0$ on that input. We define the cost function, $Cost$, of the buggy ML model with accuracy $a$ and bias value $f$ on the input. As AutoML tries different hyperparameter configurations to fix the model, the values of $a$ and $f$ may change over time. The upper bound of the cost function is defined as Equations \ref{eq:upperbound} and \ref{eq:tempBound}:
 \begin{equation}
Cost(M_{\lambda,c}, \mathcal{D}(z)) < \beta * f_0 + (1 - \beta) * (1 -  a_0)
\label{eq:upperbound}
\end{equation}	
\begin{equation}
\Leftrightarrow  \beta * f + (1 - \beta) * (1 -  a) < \beta * f_0 + (1 - \beta) * (1 -  a_0)
\label{eq:tempBound}
\end{equation}	
The upper bound of the cost function is defined with the goal of repairing a buggy model so that its performance falls within a good/win-win trade-off region of fairness and accuracy. In other words, the accuracy of the repaired model must be higher than the accuracy of the pseudo-model. The repaired model must be better than the pseudo-model in terms of the cost function's value. Since the pseudo-model has zero bias ($f_0 = 0$), the upper bound of the cost function is defined as follows (Equation \ref{eq:precondition1}):
\begin{equation}
\beta * f + (1 - \beta) * (1 -  a) < (1 - \beta) * (1 -  a_0) \\
\label{eq:precondition1}
\end{equation}	

\subsubsection{Lower bound of $\beta$} In this work, we desire to optimize the value of $\beta$ in order to minimize bias as much as possible. The cost function used by \fairautoml is designed to balance accuracy and fairness, and increasing $\beta$ will place more emphasis on reducing bias. However, simply setting $\beta$ to its highest possible value is not a viable option, as it may lead to low predictive accuracy and overfitting. We cannot accept models with poor predictive accuracy regardless of their low bias~\cite{steuer1983interactive,hakanen2017using}. To overcome this challenge, we aim to find the lower bound of $\beta$, which can be done based on the upper bound of the cost function. From Equation \ref{eq:precondition1}, we get:
\begin{equation}
\beta < \frac{a - a_0}{a - a_0 + f} 
\label{eq:precondition2}
\end{equation}	
However, if the value of $\beta$ is smaller than $\frac{a - a_0}{a - a_0 + f}$, the optimization function $Cost$ will always meet its upper bound condition. If the value of $\beta$ always satisfies the upper bound condition of the cost function regardless of accuracy and fairness, we can obtain a better optimization function by either increasing accuracy or decreasing bias. In this case, we cannot guide AutoML to produce a lower bias. Therefore, to guide AutoML produces an output with improved fairness, we set a lower bound for $\beta$ as Equation \ref{eq:lowerbound}:
\begin{equation}
	\beta \geq \frac{a - a_0}{a - a_0 + f}
    \label{eq:lowerbound}
\end{equation}

The intuition being that our method aims to increase the chance for AutoML to achieve better fairness. However, by setting $\beta < \frac{a - a_0}{a - a_0 + f}$ and $a > a_0$ (we aim to find a model which has better accuracy than the pseudo-model), any value of bias (f) can satisfy upper bound condition of the cost function, which lower chance to obtain fairer models of AutoML. To increase this chance, we set $\beta \geq \frac{a - a_0}{a - a_0 + f}$ and $a > a_0$. In this case, AutoML need to find better models that has lower bias to satisfy Equation \ref{eq:precondition1}. In other words, this lower bound condition indirectly forces bayesian optimization to search for lower bias models. 


\subsubsection{$\beta$ estimation} The final step is estimating the value of $\beta$ based on its lower bound condition. Suppose that the buggy model achieves an accuracy of $a_1$ and a bias value of $f_1$ on that input. From the begining, we have: $a = a_1$ and $f = f_1$. In that time, the lower bound of $\beta$ is $ L = \frac{a_1 - a_0}{a_1 - a_0 + f_1}$, so we have: 
\begin{equation}
    \beta = L + k, k \in [0, 1 - L]
\end{equation}
We present a greedy algorithm for estimating the value of $\beta$, which is detailed in Algorithm \ref{algorithm:greedyWeight}. Given a dataset $D$ with a protected attribute $z$ and a buggy model $M$ (Line 1), we start by measuring the lower bound of $\beta$. Next, we run \fairautoml on the input under time constraint \emph{t} with a value of $\beta$ set to $\frac{a_1 - a_0}{a_1 - a_0 + f_1}$ (Line 2-8). As the algorithm searches, whenever \fairautoml finds a candidate model that meets the condition $Cost < Cost_0$ (Lines 10-12), the value of $\beta$ is slightly increased by $\alpha$ (Line 10-12). If after N tries, \fairautoml cannot find a model that satisfies the condition, the final value of $\beta$ is set to $\beta$ = $\beta$ - $\alpha$ for the remaining search time to prevent overfitting from an excessively high value of $\beta$ (Lines 13-15). The algorithm returns the best model found (Line 16).
\setlength{\textfloatsep}{0pt}
\begin{algorithm}[t]
	\caption{Greedy Weight Identifier}
	\label{algorithm:greedyWeight}
	\begin{algorithmic}[1]
	    \State \textbf{Input:} a dataset $D$ with protected attribute $z$, buggy model $M$ hyperparametered by $\lambda$, the increment value $\alpha$, the searching time $t$ and the threshold N
		\State $\beta$ = $\frac{a_1 - a_0}{a_1 - a_0 + f_1 }$
		\State $Cost(M_{\lambda,c}, \mathcal{D}(z)) = \beta * f + (1 - \beta) * (1 -  a)$
		\State $Cost_0(M_{\lambda,c}, \mathcal{D}(z)) = \ (1 - \beta) * (1 -  a_0)$
		\State count = 0
		\State checker = False
		\While {$t$}
		\State  $M_{\lambda*,c*} = \underset{\mathcal{\lambda} \in \Lambda}{\arg \min}Cost(M_{\lambda,c}, \mathcal{D}(z))$
		\State count =  count + 1
		\If {$Cost(M_{\lambda,c}, \mathcal{D}(z)) < Cost_0(M_{\lambda,c}, \mathcal{D}(z))$}
		    \If  {checker = False}
        		\State $\beta$ = $\beta$ + $\alpha$
        		\State count = 0
        	\EndIf
		\EndIf
		\If {$count \geq N$ and checker = False}
		\State $\beta$ = $\beta$ - $\alpha$ 
		\State checker = True
		\EndIf
		\EndWhile
		\State \Return $M_\lambda*$
	\end{algorithmic}
\end{algorithm}

\subsection{Search Space Pruning for Efficient Bias Elimination}
We propose a solution to speed up the Bayesian optimization process in \fairautoml by implementing search space pruning. This technique takes advantage of data characteristics to automatically reduce the size of the search space in AutoML, thus improving its efficiency. Our approach includes two phases: the offline phase and the online phase. The offline phase trains a set of inputs multiple times to gather a collection of hyperparameters and complementary components for each input, forming a pre-built search space. In the online phase, when a new input is encountered, it is matched against the inputs stored in our database to find a matching pre-built search space, which is then utilized to repair the buggy model. This approach effectively replaces the original search space of \fairautoml, making the Bayesian optimization process much faster. Search space pruning has already been successfully applied before~\cite{cambronero2020ams, feurer2020auto}; however, this is the first application of data characteristics to prune the search space for fairness-aware AutoML.

\begin{algorithm}[t]
	\caption{Database Building}
	\label{algorithm:database}
	\begin{algorithmic}[1]
	    \State \textbf{Input:} a dataset $D$ with protected attribute $z$, a model $M$ with default hyperparameters $\lambda$. Running time t.
	    \State d = $\emptyset$
	    \State dev = 1
	    \State database = \{\}
	    \State space = \{\}
	    \State count = 0
	    \While {count $\leq$ n}
	        \State count = count + 1
    		\While {t}
    		\State $M_\lambda* = {\arg \min}Cost(M_\lambda, \mathcal{D}(z))$
    	    \State $d$ = $d$ $\cup$ $M_\lambda*$
		    \EndWhile
    		\State kBestPipelines = top_k(d)
    		\State mBestComponents = top_m(kBestPipelines)
    		\For {model $\in$ kBestPipelines}
        		\For {para $\in$ model}
        		\State space[para] = space[para]$\cup$[para.val]
        		\EndFor
    		\EndFor
    	\EndWhile
		\For {para $\in$ space}
		\If{para is numerical}
		\State no_outliers = $\emptyset$
		\For {i $\in$ space[para]}
		    \If {$|i-\overline{space[para]}| < dev*\sigma{(space[para])}$}
            \State no_outliers = no_outliers $\cup$ space[para][i]
            \EndIf
        \EndFor
        \EndIf
        
        \State space[para] = [min(no_outliers), max(no_outliers)]
		\EndFor
		\State database[input] = (space, mBestComponents)
		\State \Return database
	\end{algorithmic}
\end{algorithm}
\subsubsection{Offline Phase}
This phase constructs a set of search spaces for \fairautoml based on different inputs. It is important to note that the input format in the offline phase must match that of the online phase, which includes a dataset with a protected attribute and a ML model. This ensures that the pre-built search spaces created in the offline phase can be effectively utilized in the online phase.

\paragraph{Input} In the offline phase, we collect a set of inputs to build search spaces for \fairautoml. The inputs are obtained as follows. Firstly, we mine machine learning datasets from \openml, considering only the 3425 active datasets that have been verified to work properly. Secondly, to ensure that the mined datasets are relevant to the fairness problem, we only collect datasets that contain at least one of the following attributes: \emph{age}, \emph{sex}, \emph{race}~\cite{chakraborty2019software}. In total, we collected 231 fairness datasets. Thirdly, for each mined dataset, we use all available protected attributes. For example, when dealing with datasets that contain multiple protected attributes, such as the \emph{Adult} dataset that includes \emph{sex} and \emph{race} as protected attributes, we treat them as distinct inputs for the dataset. Finally, we use the default values for the hyperparameters of the input ML model in the offline phase, as we do not know the specific values that will be used in the online phase. 


\paragraph{Database building} To build a pre-defined search space database, we use the algorithm outlined in Algorithm \ref{algorithm:database} to obtain a pre-built search space for each collected input in order to fix the buggy model. This process involves training a fairness dataset with a specific protected attribute and ML model multiple times using \fairautoml, collecting the top $k$ best pipelines found, and extracting parameters from these pipelines. In particular, we use \fairautoml to train the fairness dataset with a specific protected attribute and a ML model for $n$ iterations (Line 7-11). We then gather the top $k$ best pipelines, including a classifier and complementary components, found by \fairautoml according to the optimization function's value (Line 12). This results in $k*n$ total pipelines. From these pipelines, we extract and store the m most frequently used complementary components in the database (Line 13). For each classifier parameter, we also store its value (Lines 14-16). This results in k$*$n values being stored for each hyperparameter. If a hyperparameter is categorical and its values are sampled from a set of different values, we store all its unique values in the database. If a hyperparameter is numerical and its values are sampled from a uniform distribution, we remove any outliers and store the range of values from the minimum to the maximum in the database (Lines 17-23). After this process, we have collected the pre-built search space for the input (Lines 24-25). We believe that two similar inputs may have similar buggy models and fixes, so the pre-built search space is built based on the best models found by \fairautoml from similar inputs, making it a reliable solution for fixing buggy models.
\begin{algorithm}[t]
	\caption{Input Matching}
	\label{algorithm:input}
	\begin{algorithmic}[1]
	    \State \textbf{Input:} a input dataset $D$ with the protected attribute $z$, the number of data points $p$, the number of features $f$, lower bound $L$, a buggy model $M$, and a database.
	    \State dist = \{\}
	    \For{$d_i$ in database}
	    \State dist[$d_i$] = $|f_i - f| + |p_i - p|$
	    \EndFor
	    \State similarDataset = min(dist, key=dist.get)
	    \State dist = \{\}
	    \For{$z_i$ in similarDataset}
	  	\State dist[$d_i$] = $|L_i - L|$
	  	\EndFor
	  	\State similarAttribute = min(dist, key=dist.get)
	  	\State similarModel = M with default parameter
	    \State \Return similarDataset, similarAttribute, similarModel
	\end{algorithmic}
\end{algorithm}
\subsubsection{Online Phase}
This phase utilizes a pre-built search space from the database to fix a buggy model for a given dataset by replacing the original search space with the pre-built one.

\paragraph{Search space pruning} Our approach of search space pruning in \fairautoml improves the bug fixing performance by reducing the size of the hyperparameter tuning space. Algorithm \ref{algorithm:input} is used to match the input dataset, protected attribute, and ML model to the most similar input in the database. Firstly, data characteristics such as the number of data points and features are used to match the new dataset with the most similar one in the database~\cite{feurer2020auto}. L1 distance is computed between the new dataset and each mined dataset in the space of data characteristics to determine the closest match. We consider that the most similar dataset to the new dataset is the nearest one (Line 2-5). Secondly, we compute the lower bound $L = \frac{a_1 - a_0}{a_1 - a_0 + f_1}$ of $\beta$ of the new input. We then estimate the lower bound of $\beta$ of all the protected attributes of the matched dataset and select the attribute whose lower bound is closest to $L$ (Line 6-9). Lastly, two similar inputs must use the same ML algorithm (Line 10). The matching process is carried out in the order of dataset matching, protected attribute matching, and ML algorithm matching. The pre-built search space of the similar input is then used as the new search space for the new input.

\section{Evaluation}
\label{sec:eval}

\begin{table*}[t]
\centering
\caption{Trade-off assessment results of \fairautoml, \as, and mitigation techniques 
}
\setlength{\tabcolsep}{0.75pt}
 \footnotesize
\begin{tabular}{|c|c|c|c|c|c|c|c|c|c|c|c|c|c|c|c|c|c|c|c|c|c|c|c|c|c|c|c|}
\cline{1-14} \cline{16-28}
                                  & Model                  & Metric & T1                                    & T2                                    & T3                                    & T4                                    & AS             & R                                     & DIR            & PML                                   & EO                           & FAX                                   & ROC                          &  & Model                  & Metric & T1                                    & T2                                    & T3                                    & T4                                    & AS                                    & R                                     & DIR            & PML                                   & EO                           & FAX                           & ROC                          \\ \cline{1-14} \cline{16-28} 
                                  &                        & Acc    & 0.010                                 & 0.001                                 & 0.005                                 & -0.003                                & 0.016          & 0.009                                 & 0.004          & 0.008                                 & -0.006                       & 0.002                                 & -0.047                       &  &                        & Acc    & -0.013                                & -0.022                                & -0.009                                & -0.032                                & -0.001                                & -0.056                                & -0.055         & -0.054                                & -0.057                       & -0.057                        & -0.061                       \\
\end{tabular}
\label{tbl:regions}
\centering
\footnotesize \*  Each cell shows the accuracy/bias difference between the original and repaired models. For accuracy, accuracy difference = new accuracy - old accuracy. For bias (DI, SPD, EOD, AOD), bias difference = old bias - new bias. Thus, a positive value indicates an improvement in bias/accuracy in the repaired model compared to the original and vice versa. For bias, if a method falls into either the good region (regular numbers) or the win-win region (\textbf{bold numbers}), the bias difference value will be provided. If it falls into any other region, the region type will be indicated. The values highlighted in {\color[HTML]{0070C0}blue} denote the most effective bug fixing method. The data from this table is divided and analyzed in depth in Tables \ref{tbl:percentage}, \ref{tbl:summary}, \ref{tbl:sp}.
\vspace{-1em}
\end{table*}

In this section, we describe the design of the experiments to evaluate the efficient of \fairautoml. We first pose research questions and discuss the experimental details. Then, we answer research questions regarding the efficiency and adaptability of \fairautoml.

\textbf{RQ1: Is \fairautoml effective in fixing fairness bugs?} 
 To answer this question, we quantify the number of fairness bugs that \fairautoml is able to repair compared to existing methods, allowing us to assess the capability of an AutoML system in fixing fairness issues.

\textbf{RQ2: Is \fairautoml more adaptable than existing bias mitigation techniques?} The adaptability of a bias mitigation technique indicates its performance across a diverse range of datasets/metrics. So,  we analyze the effectiveness of
\fairautoml and existing bias mitigation techniques on different
dataset/metrics to assess the adaptability of an AutoML system on fix fairness bugs.

\textbf{RQ3: Are dynamic optimization function and search space pruning effective in fixing fairness bugs?} To answer this question, we assess the performance of \as, both with and without the dynamic optimization function and search space pruning, to demonstrate the impact of each proposed approach.


\subsection{Experiment}
\subsubsection{Benchmarks}
We evaluated our method using real-world fairness bugs sourced from a recent empirical study~\cite{biswas20machine}, with our benchmark consisting of 16 models collected from Kaggle covering five distinct types: XGBoost (\textbf{XGB}), Random Forest (\textbf{RF}), Logistic Regression (\textbf{LRG}), Gradient Boosting (\textbf{GBC}), Support Vector Machine (\textbf{SVC}). We use four popular datasets for our evaluation~\cite{bogen2018help, tramer2017fairtest, udeshi2018automated}: 

The \textbf{Adult Census} (race)~\cite{adult2017kaggle} comprised of 32,561 observations and 12 features that capture the financial information of individuals from the 1994 U.S. census. The objective is to predict whether an individual earns an annual income greater than 50K.

The \textbf{Bank Marketing} (age)~\cite{bank2017kaggle} has 41,188 data points with 20 features including information on direct marketing campaigns of a Portuguese banking institution. The classification task aims to identify whether the client will subscribe to a term deposit. 

The \textbf{German Credit} (sex)~\cite{german2017kaggle} has 1000 observations with 21 features containing credit information to predict good or bad credit. 

The \textbf{Titanic} (sex)~\cite{titanic2017kaggle} has 891 data points with 10 features containing individual information of Titanic passengers. The dataset is used to predict who survived the Titanic shipwreck.

\begin{table*}[t]
\centering
\caption{Proportion of \fairautoml, \as, and mitigation techniques that fall into each mitigation region}
\setlength{\tabcolsep}{0.47pt}
\footnotesize
\selectfont

\begin{tabular}{|c|cccccccccccccccccccc|l|cccccccccccccccccccc|}
\cline{1-21} \cline{23-42}
                           & \multicolumn{20}{c|}{\scriptsize {Fairness Metric}}                                                                                                                                                                                                                                                                                                                                                                                                                                                                                                                                                                                                                                      &  & \multicolumn{20}{c|}{\scriptsize {Dataset}}                                                                                                                                                                                                                                                                                                                                                                                                                                                                                                                                                                                                                                                     \\
\end{tabular}

\label{tbl:percentage}
\footnotesize \* The proportions in this table are determined based on the data presented in Table \ref{tbl:regions}:  proportion for fairness metric = \# buggy cases of a metric fall into a region / \# buggy cases of that metric, proportion for dataset = \# buggy cases fall of a dataset into a region / \# buggy cases of that dataset.
\vspace{-1em}
\end{table*}

\subsubsection{Evaluated Learning Techniques}
We examined the performance of \fairautoml and other supervised learning methods addressing discrimination in binary classification including all three types of bias mitigation techniques and Auto-ML techniques. 

\paragraph{Bias mitigation methods} We investigate all three types of bias mitigation methods: pre-processing, in-processing, post-processing. We select widely-studied bias mitigation methods for each category: 
\begin{itemize}
\item The \textbf{pre-processing} includes \emph{Reweighing} (\textbf{R})~\cite{kamiran2012data}, \emph{Disparate Impact Remover} (\textbf{DIR})~\cite{feldman2015certifying}.
\item The \textbf{in-processing} includes \emph{Parfait-ML} (\textbf{PML})~\cite{tizpaz2022fairness}.
\item The \textbf{post-processing} includes \emph{Equalized Odds} (\textbf{EO})~\cite{hardt2016equality}, \emph{FaX-AI} (\textbf{FAX})~\cite{grabowicz2022marrying}, \emph{Reject Option Classification} (\textbf{ROC})~\cite{kamiran2012decision}.
\end{itemize}

\paragraph{Auto-Sklearn} We explore the efficiency of \as (\textbf{AS})~\cite{feurer2015efficient} on mitigating bias in unfair model. Although, \as does not seek to decrease bias, we compare its performance with \fairautoml to demonstrate the efficient of our techniques in guiding Auto-ML to repair fairness bugs. 

\paragraph{Fair-AutoML} We create 4 versions of \fairautoml in this evaluation representing for \fairautoml with different cost functions:
\begin{itemize}
\item \textbf{T1} uses  $\beta * DI + (1 - \beta) * (1 - accuracy)$ as a cost function.

\item \textbf{T2} uses  $\beta*SPD + (1 - \beta)*(1 - accuracy)$ as a cost function.

\item \textbf{T3} uses  $\beta*EOD + (1 - \beta)*(1 - accuracy)$ as a cost function.

\item \textbf{T4} uses  $\beta*AOD + (1 - \beta)*(1 - accuracy)$ as a cost function.
\end{itemize}
\subsubsection{Experimental Configuration}
Experiments were conducted using Python 3.6 on Intel Skylake 6140 processors. \fairautoml leverages the capabilities of \as~\cite{feurer2015efficient}, taking advantage of its automatic optimization of the best ML model for a given dataset. We tailored \as to better fit our method in two ways: (1) its search space was restricted to the type of the faulty classifier - for example, if the faulty classifier is Random Forest, \as will only optimize the hyperparameters and identify complementary components for that specific classifier. (2) The faulty model was set as the default model for \as. These modifications are features of \as that we utilized.



\paragraph{\textbf{Methodology Configuration}}
We selected an increment value of $\alpha$ for $\beta$ of 0.05 to balance the time between $\beta$ search and model fixing processes. The user can opt for a more accurate value of $\beta$ by decreasing the increment value and using a longer search time. To conduct search space pruning, we ran \fairautoml 10 times (n) with a 1-hour search time (t) to gather the best ML pipelines \cite{biswas22art}. From each run, we collected the top 10 pipelines (k), resulting in 100 models per input. This pre-built search space includes a set of hyperparameters and the top 3 most frequently used complementary components (m). We have explored other parameter settings, but these have proven to provide optimal results.



\paragraph{\textbf{Evaluation Configuration}}
We evaluate each tool on each buggy scenario 10 times using a random re-split of the data based on a 7:3 train-test split ratio~\cite{hort2021fairea}. The runtime for each run of \fairautoml and \as is approximately one hour~\cite{feurer2015efficient, feurer2020auto}. The mean performance of each method is calculated as the average of the 10 runs, which is a commonly used practice in the fairness literature~\cite{bellamy2018ai, biswas20machine, celis2019classification}. Our evaluation targets fixing 16 buggy models for 4 fairness metrics, resulting in a total of 64 buggy cases. 





\subsection{Effectiveness (RQ1)}

We evaluate the effectiveness of \fairautoml by comparing it with \as and existing bias mitigation techniques based on \fairea baseline. The comparisons are based on the following rules:

\begin{itemize}
\item \textbf{Rule 1}: A model is considered successfully repaired when its post-mitigation mean accuracy and fairness falls into win-win/good trade-off regions.
\item \textbf{Rule 2}: A model that falls in the win-win region is always better than one falling into any other region.
\item \textbf{Rule 3}: If two models are in the same trade-off region, the one with lower bias is preferred. 
\end{itemize}

Our comparison rules for bug-fixing performance were established based on Fairea and our evaluations. Firstly, we define a successful bug fix as a fixed model that falls within the win-win or good trade-off regions, as these regions demonstrate improved fairness-accuracy trade-offs compared to the baseline in Fairea. Secondly, when comparing successfully fixed models in different trade-off regions (win-win versus good), we consider the win-win models to be superior as they offer improved fairness and accuracy. Lastly, for models that fall within the same trade-off region, the one with lower bias is deemed to be better, as our goal is to fix unfair models. Our evaluations then consider two aspects of the bug-fixing performance: the number of successful bug fixes and the number of times a bias mitigation method outperforms others.

\subsubsection{Is \fairautoml effective in fixing fairness bugs?} The results presented in Table \ref{tbl:summary} show that \fairautoml was effective in resolving 60 out of 64 (94\%) fairness bugs, while \as only fixed 28 out of 64 (44\%) and bias mitigation techniques resolved up to 44 out of 64 (69\%). This indicates that \as alone was not effective in reducing bias, however, our methods were successful in enhancing AutoML to repair fairness bugs. Moreover, \fairautoml was able to repair more cases than other bias mitigation techniques, which often resulted in lower accuracy for lower bias. This highlights the effectiveness of our approaches in guiding AutoML towards repairing models for better trade-off between fairness and accuracy compared to the Fairea baseline. 
\begin{table}
\centering
\caption{Fair-AutoML (FA) vs bias mitigation methods in fixing fairness bugs}
\setlength{\tabcolsep}{6pt}
\footnotesize
\begin{tabular}{|l|r|r|r|r|r|r|r|r|}
\hline
                           & FA & AS & R  & DIR  & PML  & EO  & FAX  & ROC  \\ \hline
\# bugs fixed & 60          & 28 & 35 & 30 & 36 & 37 & 44 & 23     \\ \hline
\# best models   & 19          & 4  & 9  & 0  & 10  & 9 & 9  & 3      \\ \hline
\end{tabular}
\label{tbl:summary}
\footnotesize \* The results in this table are derived from the data presented in Table \ref{tbl:regions}. The row \emph{\# bugs fixed} indicates the number of cases where the technique falls into either the win-win or good trade-off region. The row \emph{\# best models} represents the number of instances where a bias mitigation technique outperforms all other methods. 
\end{table} 
 \subsubsection{Does \fairautoml outperform bias reduction techniques?}
\fairautoml demonstrated superior performance in fixing fairness bugs compared to other bias mitigation techniques. The results presented in Table \ref{tbl:summary} indicate that 63 out of 64 buggy cases were fixed by \fairautoml, \as, or bias mitigation techniques. Among the repaired buggy cases, \fairautoml outperformed other techniques 19 times (30\%). On the other hand, \as outperformed \fairautoml and bias mitigation techniques only 4 times (6\%), and bias mitigation techniques outperformed other techniques 10 times at most (16\%). This highlights that \fairautoml is often more effective in improving fairness and accuracy simultaneously or reducing more bias than other bias mitigation techniques.


\subsection{Adaptability (RQ2)}
To assess the adaptability of \fairautoml, we measure the proportions of each evaluated tools that fall into each fairness-accuracy trade-off region in different categories: fairness metric and dataset (Table \ref{tbl:percentage}). To further evaluate the adaptability of \fairautoml, instead of using our prepared models and datasets, we used the benchmark~\cite{tizpaz2022fairnessartifact} of Parfait-ML to evaluate \fairautoml. Particularly, we evaluate \fairautoml and Parfait-ML on three different ML models (Decision Tree, Logistic Regression, Random Forest) on two datasets (Adult Census and COMPAS) (Table \ref{tbl:pml} and Figure \ref{fig:pml}).


\subsubsection{Is \fairautoml more adaptable than existing bias mitigation techniques and \as?}
	

Table \ref{tbl:percentage} shows \fairautoml demonstrates exceptional repair capabilities across various datasets and fairness metrics, with a high rate of success in fixing buggy models. For example, in the Adult Census, Bank Marketing, German Credit, and Titanic datasets, \fairautoml (T4) repaired 100\%, 82\%, 94\%, and 94\% of the models, respectively. Similarly, in the DI, SPD, EOD, and AOD fairness metrics, \fairautoml (T4) achieved repair rates of 100\%, 94\%, 82\%, and 94\%. On the other hand, bias mitigation methods often show inconsistent results. For instance, Equalized Odds repaired all buggy cases in \emph{Adult Census} but none in \emph{Bank Marketing}. In fact, our methods effectively guides AutoML in hyperparameter tuning to reduce bias, leading to superior repair performance across different datasets and metrics.

\subsubsection{Is Fair-AutoML effective in fixing fairness bugs on other bias mitigation methods benchmark?}
\begin{figure}
	\includegraphics[width=0.48\textwidth]{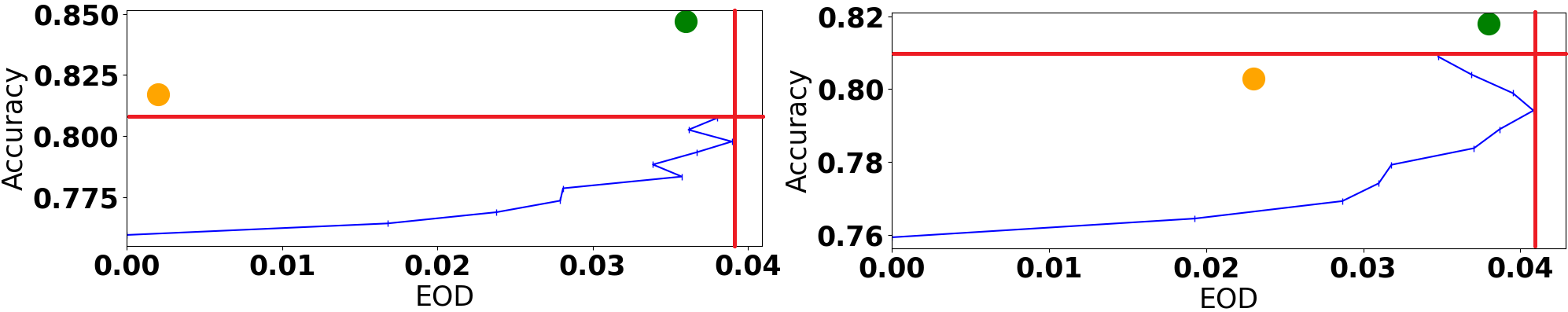}
	\vspace{-2em}
	\caption{Accuracy and fairness achieved by Fair-AutoML (green circle) and Pafait-ML (orange circle) with Decision Tree (left) and logistic regression (right) on Adult dataset (Pafait-ML’s benchmark). The blue line shows the Fairea baseline and red lines define the trade-off regions.}
	\label{fig:pml}
\end{figure}

Based on evaluation of Parfait-ML~\cite{tizpaz2022fairness}, we only use accuracy and EOD as evaluation metrics for this evaluation. To make a fair comparison with Parfait-ML, we utilize the version of \fairautoml that incorporates EOD and accuracy as its cost function (T3). The results are displayed in Table \ref{tbl:pml}, showcasing the accuracy and bias (EOD) achieved by both \fairautoml (T3) and Parfait-ML in Parfait-ML's benchmark. The table showcases the actual results of the repaired models, rather than the difference in accuracy/fairness between the original and repaired models. Upon inspection, the results for the COMPAS dataset for both \fairautoml and Parfait-ML are similar. However, for the Adult dataset, some differences arise. For instance, with the Random Forest classifier, \fairautoml performs better than Parfait-ML in both accuracy and EOD. With the Logistic Regression classifier, \fairautoml achieved a higher accuracy but higher bias compared to Parfait-ML. Nevertheless, \fairautoml falls into the win-win trade-off region, while
Parfait-ML only falls into good trade-off region (Figure \ref{fig:pml}). With the Decision Tree classifier, both \fairautoml and Parfait-ML fall into the win-win trade-off region (Figure \ref{fig:pml}); however, Parfait-ML performed better since it has lower bias. These results highlights the generalization capability of \fairautoml to repair various datasets and ML models.

\begin{table}
\centering
\caption{Accuracy and fairness achieved by Fair-AutoML and Pafait-ML on Pafait-ML's benchmark}
\setlength{\tabcolsep}{1.45pt}
\footnotesize
\begin{tabular}{|cl|cccc|cccc|cccc|}
\hline
\multicolumn{2}{|c|}{\multirow{3}{*}{Data}} & \multicolumn{4}{c|}{Decision Tree}                                                                      & \multicolumn{4}{c|}{Logistic Regression}                                                                     & \multicolumn{4}{c|}{Random Forest}                                                                      \\ \cline{3-14} 
\multicolumn{2}{|c|}{}                      & \multicolumn{2}{c|}{T3}                                 & \multicolumn{2}{c|}{PML}           & \multicolumn{2}{c|}{T3}                                 & \multicolumn{2}{c|}{PML}           & \multicolumn{2}{c|}{T3}                                 & \multicolumn{2}{c|}{PML}           \\ \cline{3-14} 
\multicolumn{2}{|c|}{}                      & \multicolumn{1}{c|}{Acc}   & \multicolumn{1}{c|}{EOD}   & \multicolumn{1}{c|}{Acc}   & EOD   & \multicolumn{1}{c|}{Acc}   & \multicolumn{1}{c|}{EOD}   & \multicolumn{1}{c|}{Acc}   & EOD   & \multicolumn{1}{c|}{Acc}   & \multicolumn{1}{c|}{EOD}   & \multicolumn{1}{c|}{Acc}   & EOD   \\ \hline
\multicolumn{2}{|c|}{Adult}                & \multicolumn{1}{c|}{0.847} & \multicolumn{1}{c|}{0.036} & \multicolumn{1}{c|}{0.817} & 0.002 & \multicolumn{1}{c|}{0.818} & \multicolumn{1}{c|}{0.038} & \multicolumn{1}{c|}{0.803} & 0.023 & \multicolumn{1}{c|}{0.851} & \multicolumn{1}{c|}{0.032} & \multicolumn{1}{c|}{0.843} & 0.039 \\ \hline
\multicolumn{2}{|c|}{Compas}                & \multicolumn{1}{c|}{0.969} & \multicolumn{1}{c|}{0.000} & \multicolumn{1}{c|}{0.970} & 0.000 & \multicolumn{1}{c|}{0.970} & \multicolumn{1}{c|}{0.000} & \multicolumn{1}{c|}{0.968} & 0.000 & \multicolumn{1}{c|}{0.970} & \multicolumn{1}{c|}{0.000} & \multicolumn{1}{c|}{0.970} & 0.000 \\ \hline
\end{tabular}
\label{tbl:pml}

\end{table}

\subsection{Ablation Study (RQ3)}
We create an ablation study to observe the efficiency of the dynamic optimization function and the search space pruning separately. The ablation study compares the performance of the following tools:   
\begin{itemize}
\item \as (\emph{AS}) represents AutoML.
\item \fairautoml version 1 (\emph{FAv1}) represents AutoML + dynamic optimization function.
\item \fairautoml version  2 (\emph{FAv2}) represents AutoML + dynamic optimization function + search space pruning.
\end{itemize}
 
To evaluate the efficiency of the dynamic
optimization function, we compare the performance of \emph{FAv1} with \as. We compare \emph{FAv1} with \emph{FAv2} to observe the efficiency of the search space pruning approach. The complete result is shown in Table \ref{tbl:sp}. Notice that we use \fairautoml to optimize different fairness metrics; thus, we only consider the metric that each tool tries to optimize. For instance, the results of Random Forest on Adult dataset in the Table \ref{tbl:sp} shows that achieved scores of 0.096 for DI, 0.014 for SPD, 0.024 for EOD, and 0.035 for AOD. This result means that T1 achieves 0.096 for DI, T2 achieves 0.014 for SPD, T3 achieves 0.024 for EOD, T4 achieves 0.035 for AOD. The evaluation only considers cases where the tools successfully repair the bug. The same rules described in RQ1 is applied in this evaluation.

\begin{table}[t]
\centering
\caption{Trade-off assessment results of \as, \emph{FAv1}, and \emph{FAv2}}
\setlength{\tabcolsep}{3.8pt}
\footnotesize
\begin{tabular}{|c|c|c|c|c|c|c|c|c|c|}
\hline
\textbf{}                 &  Metric & Model         & AS                  & FAv1                         & FAv2                         & Model         & AS                           & FAv1                         & FAv2                         \\ \hline
                          & DI                                                  &                        & 0.058                        & {\color[HTML]{ED7D31} 0.119}                                       & {\color[HTML]{0070C0}\textbf{0.096}}                           &                        & 0.04                                  & {\color[HTML]{ED7D31} 0.094}                                       & {\color[HTML]{0070C0} 0.19}                                     \\ \cline{2-2} \cline{4-6} \cline{8-10} 
                          & SPD                                                 &                        & 0.016                        & {\color[HTML]{ED7D31} 0.053}                                       & {\color[HTML]{0070C0} \textbf{0.024}}                           &                        & 0.005                                 & {\color[HTML]{ED7D31} 0.041}                                       & {\color[HTML]{0070C0} 0.075}                                    \\ \cline{2-2} \cline{4-6} \cline{8-10} 
                          & EOD                                                 &                        & 0.008                        & {\color[HTML]{0070C0} \textbf{0.015}}                              & {\color[HTML]{ED7D31} \textbf{0.014}}                           &                        & lose                                  & {\color[HTML]{ED7D31} 0.043}                                       & {\color[HTML]{0070C0} 0.044}                                    \\ \cline{2-2} \cline{4-6} \cline{8-10} 
                          & AOD                                                 & \multirow{-4}{*}{RF}   & 0.021                        & {\color[HTML]{0070C0} \textbf{0.025}}                              & {\color[HTML]{ED7D31} 0.035}                                    & \multirow{-4}{*}{LRG}  & 0.039                                 & {\color[HTML]{ED7D31} 0.078}                                       & {\color[HTML]{0070C0} 0.089}                                    \\ \cline{2-10} 
                          & DI                                                  &                        & 0.004                        & {\color[HTML]{ED7D31} 0.136}                                       & {\color[HTML]{0070C0} 0.183}                                    &                        & inv                                   & bad                                                                & {\color[HTML]{0070C0} 0.124}                                    \\ \cline{2-2} \cline{4-6} \cline{8-10} 
                          & SPD                                                 &                        & 0.003                        & {\color[HTML]{ED7D31} 0.046}                                       & {\color[HTML]{0070C0} 0.074}                                    &                        & inv                                   & bad                                                                & {\color[HTML]{0070C0} 0.055}                                    \\ \cline{2-2} \cline{4-6} \cline{8-10} 
                          & EOD                                                 &                        & lose                         & {\color[HTML]{ED7D31} 0.015}                                       & {\color[HTML]{0070C0} 0.037}                                    &                        & inv                                   & {\color[HTML]{0070C0} 0.02}                                        & {\color[HTML]{ED7D31} 0.013}                                    \\ \cline{2-2} \cline{4-6} \cline{8-10} 
\multirow{-8}{*}{\rotatebox[origin=c]{90}{Adult Census}} & AOD                                                 & \multirow{-4}{*}{XGB}  & 0.017                        & {\color[HTML]{ED7D31} 0.049}                                       & {\color[HTML]{0070C0} 0.053}                                    & \multirow{-4}{*}{GBC}  & {\color[HTML]{0070C0} \textbf{0.018}} & inv                                                                & {\color[HTML]{ED7D31} 0.05}                                     \\ \hline
                          & DI                                                  &                        & 0.000                            & {\color[HTML]{0070C0} 0.663}                                       & {\color[HTML]{ED7D31} 0.103}                                    &                        & lose                                  & inv                                                                & {\color[HTML]{0070C0} 0.158}                                    \\ \cline{2-2} \cline{4-6} \cline{8-10} 
                          & SPD                                                 &                        & lose                         & {\color[HTML]{0070C0} 0.026}                                       & bad                                                             &                        & lose                                  & inv                                                                & {\color[HTML]{0070C0} 0.051}                                    \\ \cline{2-2} \cline{4-6} \cline{8-10} 
                          & EOD                                                 &                        & lose                         & inv                                                                & lose                                                            &                        & lose                                  & lose                                                               & {\color[HTML]{0070C0} \textbf{0.003}}                           \\ \cline{2-2} \cline{4-6} \cline{8-10} 
                          & AOD                                                 & \multirow{-4}{*}{RF}   & 0.016                        & {\color[HTML]{0070C0} \textbf{0.004}}                              & {\color[HTML]{ED7D31} 0.032}                                    & \multirow{-4}{*}{XGB2}  & lose                                  & {\color[HTML]{ED7D31} 0.003}                                       & {\color[HTML]{0070C0} \textbf{0.027}}                           \\ \cline{2-10} 
                          & DI                                                  &                        & lose                         & inv                                                                & {\color[HTML]{0070C0} 0.098}                                    &                        & lose                                  & {\color[HTML]{0070C0} 0.014}                                       & {\color[HTML]{ED7D31} 0.01}                                     \\ \cline{2-2} \cline{4-6} \cline{8-10} 
                          & SPD                                                 &                        & lose                         & {\color[HTML]{ED7D31} 0.018}                                       & {\color[HTML]{0070C0} \textbf{0.062}}                           &                        & lose                                  & lose                                                               & {\color[HTML]{0070C0} 0.028}                                    \\ \cline{2-2} \cline{4-6} \cline{8-10} 
                          & EOD                                                 &                        & lose                         & inv                                                                & {\color[HTML]{0070C0} \textbf{0.011}}                           &                        & lose                                  & lose                                                               & inv                                                             \\ \cline{2-2} \cline{4-6} \cline{8-10} 
\multirow{-8}{*}{\rotatebox[origin=c]{90}{Bank Marketing}}  & AOD                                                 & \multirow{-4}{*}{XGB1}  & lose                         & lose                                                               & {\color[HTML]{0070C0} \textbf{0.046}}                           & \multirow{-4}{*}{GBC}  & lose                                  & {\color[HTML]{ED7D31} 0.003}                                       & {\color[HTML]{0070C0} \textbf{0.025}}                           \\ \hline
                          & DI                                                  &                        & lose                         & bad                                                                & bad                                                             &                        & 0.035                                 & {\color[HTML]{0070C0} 0.127}                                       & {\color[HTML]{ED7D31} 0.109}                                    \\ \cline{2-2} \cline{4-6} \cline{8-10} 
                          & SPD                                                 &                        & lose                         & bad                                                                & {\color[HTML]{0070C0} 0.052}                                    &                        & 0.021                                 & {\color[HTML]{ED7D31} 0.078}                                       & {\color[HTML]{0070C0} 0.092}                                    \\ \cline{2-2} \cline{4-6} \cline{8-10} 
                          & EOD                                                 &                        & {\color[HTML]{ED7D31} 0.033} & bad                                                                & {\color[HTML]{0070C0} 0.045}                                    &                        & 0.068                                 & {\color[HTML]{0070C0} 0.112}                                       & {\color[HTML]{ED7D31} 0.101}                                    \\ \cline{2-2} \cline{4-6} \cline{8-10} 
                          & AOD                                                 & \multirow{-4}{*}{RF}  & lose                         & lose                                                               & lose                                                            & \multirow{-4}{*}{SVC}   & lose                                  & {\color[HTML]{0070C0} 0.032}                                       & {\color[HTML]{ED7D31} 0.027}                                    \\ \cline{2-10} 
                          & DI                                                  &                        & bad                          & bad                                                                & {\color[HTML]{0070C0} \textbf{0.07}}                            &                        & {\color[HTML]{ED7D31}0.038}                                 & inv                                                                & {\color[HTML]{0070C0} \textbf{0.112}}                           \\ \cline{2-2} \cline{4-6} \cline{8-10} 
                          & SPD                                                 &                        & bad                          & lose                                                               & {\color[HTML]{0070C0} 0.065}                                    &                        & 0.027                                 & {\color[HTML]{ED7D31} \textbf{0.012}}                              & {\color[HTML]{0070C0} \textbf{0.075}}                           \\ \cline{2-2} \cline{4-6} \cline{8-10} 
                          & EOD                                                 &                        & {\color[HTML]{ED7D31} 0.036} & lose                                                               & {\color[HTML]{0070C0} 0.073}                                    &                        & 0.066                                 & {\color[HTML]{ED7D31} \textbf{0.050}}                               & {\color[HTML]{0070C0} \textbf{0.085}}                           \\ \cline{2-2} \cline{4-6} \cline{8-10} 
\multirow{-8}{*}{\rotatebox[origin=c]{90}{German Credit}}   & AOD                                                 & \multirow{-4}{*}{XGB}  & lose                         & lose                                                               & {\color[HTML]{0070C0} 0.037}                                    & \multirow{-4}{*}{KNN}  & lose                                  & inv                                                                & {\color[HTML]{0070C0} 0.034}                                    \\ \hline
                          & DI                                                  &                        & lose                         & {\color[HTML]{ED7D31} 1.04}                                        & {\color[HTML]{0070C0} 1.549}                                    &                        & {\color[HTML]{0070C0} \textbf{0.092}} & 0.447                                                              & {\color[HTML]{ED7D31} 0.501}                                    \\ \cline{2-2} \cline{4-6} \cline{8-10} 
                          & SPD                                                 &                        & lose                         & {\color[HTML]{ED7D31} 0.525}                                       & {\color[HTML]{0070C0} 0.571}                                    &                        & inv                                   & {\color[HTML]{ED7D31} 0.273}                                       & {\color[HTML]{0070C0} 0.275}                                    \\ \cline{2-2} \cline{4-6} \cline{8-10} 
                          & EOD                                                 &                        & lose                         & {\color[HTML]{ED7D31} 0.386}                                       & {\color[HTML]{0070C0} 0.445}                                    &                        & {\color[HTML]{0070C0} \textbf{0.058}} & {\color[HTML]{ED7D31} 0.184}                                       & bad                                                             \\ \cline{2-2} \cline{4-6} \cline{8-10} 
                          & AOD                                                 & \multirow{-4}{*}{RF}   & 0.062                        & {\color[HTML]{ED7D31} 0.577}                                       & {\color[HTML]{0070C0} 0.601}                                    & \multirow{-4}{*}{GBC} & {\color[HTML]{0070C0} \textbf{0.058}} & 0.429                                                              & {\color[HTML]{ED7D31} 0.445}                                    \\ \cline{2-10} 
                          & DI                                                  &                        & 0.086                        & {\color[HTML]{0070C0} 1.062}                                       & {\color[HTML]{ED7D31} 0.743}                                    &                        & lose                                  & {\color[HTML]{ED7D31} 1.205}                                       & {\color[HTML]{0070C0} 1.364}                                    \\ \cline{2-2} \cline{4-6} \cline{8-10} 
                          & SPD                                                 &                        & 0.063                        & {\color[HTML]{0070C0} 0.594}                                       & {\color[HTML]{ED7D31} 0.552}                                    &                        & lose                                  & {\color[HTML]{ED7D31} 0.314}                                       & {\color[HTML]{0070C0} 0.542}                                    \\ \cline{2-2} \cline{4-6} \cline{8-10} 
                          & EOD                                                 &                        & lose                         & {\color[HTML]{ED7D31} 0.556}                                       & {\color[HTML]{0070C0} 0.557}                                    &                        & lose                                  & {\color[HTML]{0070C0} 0.287}                                       & {\color[HTML]{ED7D31} 0.285}                                    \\ \cline{2-2} \cline{4-6} \cline{8-10} 
\multirow{-8}{*}{\rotatebox[origin=c]{90}{Titanic}}    & AOD                                                 & \multirow{-4}{*}{LRG} & 0.101                        & {\color[HTML]{ED7D31} 0.651}                                       & {\color[HTML]{0070C0} \textbf{0.179}}                           & \multirow{-4}{*}{XGB}  & lose                                  & {\color[HTML]{ED7D31} 0.441}                                       & {\color[HTML]{0070C0} 0.524}                                    \\ \hline
\end{tabular}
\label{tbl:sp}
\centering
\footnotesize \*  The data in Table \ref{tbl:sp} is created in the same ways as Table \ref{tbl:regions}.  For each method in the
good trade-off region and win-win region (\textbf{bold number}), a trade-off measurement value is given; for
other regions the region type is displayed. The values in {\color[HTML]{0070C0}blue}, {\color[HTML]{ED7D31}orange}, and black indicate the top 1, top 2, top 3 bug fixing tools, respectively. 
\end{table}

\subsubsection{Are dynamic optimization function and search space pruning effective in fixing fairness bugs?}
From Table \ref{tbl:sp}, our results show that the dynamic optimization function approach in \fairautoml helps fix buggy models more efficiently. Comparing the performance in fixing fairness bugs, \emph{FAv1} outperforms \as 39 times, while \as outperforms \emph{FAv1} only 7 times. The search space pruning approach in \fairautoml also contributes to more efficient bug fixing, as \emph{FAv2} outperforms both \emph{FAv1} and \as 46 and 55 times respectively, while \emph{FAv1} and \as only outperform \emph{FAv2} 14 and 4 times respectively.

\section{Discussion}
\label{sec:discussion}

In this work, we bring particular attention to the fairness-accuracy tradeoff while mitigating bias in ML models. Many works in the area only optimize fairness metrics by sacrificing accuracy, and do not consider the tradeoff rigorously. However, as shown by recent work \cite{hort2021fairea}, trivial mutation methods can also achieve fairness if accuracy is compromised in different magnitudes. Therefore, a rigorous evaluation method is necessary to demonstrate that the tradeoff is beneficial. Another limitation of existing tools is not generalizing over different ML classifiers (e.g., LRG, GBC, RF, XGB), multiple fairness metrics, and dataset characteristics. To that end, we leveraged the recent progress of AutoML in the context and achieved better tradeoff than SOTA methods. 
We believe that our approach is versatile and can be applied to various ML problems. Particularly, the dynamic optimization function approach remains versatile across various datasets and models. 
Furthermore, the search space pruning approach is refined through pre-constructed database and a matching mechanism, that capitalizes on diverse datasets stored in repositories such as \openml or \kg. 

We implemented \fairautoml on top of \as to ensure its wide applicability on ML algorithms. State-of-the-art bias mitigation techniques also primarily use classic ML algorithms ~\cite{biswas20machine, biswas2021fair, chakraborty2021bias, chakraborty2020fairway, chen2022maat, tizpaz2022fairness} that are supported by \as. These models are more suitable than the DL models since the fairness critical tasks in prior works commonly use tabular datasets. 
Should one desire to explore alternative model types not directly supported by \as, they can adopt the general ML model adoption of \as~\cite{asdoc}. 

Our approach also outlines several opportunities towards leveraging AutoML and search-based software engineering to ensure fairness in new ML models that are becoming available. First, the greedy weight identifier algorithm's performance might suffer for complex models due to computational costs (Algorithm \ref{algorithm:greedyWeight}). Second, search space pruning quantitatively estimates the similarity of datasets based on data characteristics. Thus, if we do not have a dataset similar enough to the input dataset, AutoML may not perform well. To address this, we plan to regularly update our database with new datasets. 
Lastly, constructing suitable search spaces, particularly for resource-intensive methods like deep learning, could entail significant computational expenses. 
Further works are needed to maximize the versatility and effectiveness of our approach over novel fairness-critical tasks. 
One key direction is to combine \fairautoml with other bias mitigation techniques, such as integrating Fair-AutoML's model with pre-processing bias mitigation methods to enhance overall pipeline fairness. Additionally, integrating \fairautoml with ensemble learning could improve both performance and fairness by capturing a broader range of biases and patterns. These directions could significantly amplify the impact of this work, making Fair-AutoML a potent tool for promoting fairness and equity in machine learning across various domains.



\section{Threats to Validity}
\label{sec:limit}

\paragraph{Construct Validity}
The choice of evaluation metrics and existing mitigation techniques may pose a threat to our results. We mitigate this threat by employing a diverse range of metrics and mitigation methods. First,  we have used accuracy and four most recent and widely-used fairness metrics to evaluate \fairautoml and the state-of-the-art. These metrics have been commonly applied in the software engineering community~\cite{chakraborty2021bias, chakraborty2020fairway, chen2022maat, tizpaz2022fairness}. Second, we demonstrate the superiority of \fairautoml over state-of-the-art methods in different categories: pre-processing, in-processing, and post-processing, which are most advanced techniques from the SE and ML communities. For evaluating fairness and applying these mitigation algorithms except Parfait-ML~\cite{tizpaz2022fairness}, we have used AIF 360 toolkit. For evaluating Parfait-ML, we have used its original implementation. We create a baseline using the original Fairea implementation, enabling us to conduct a comprehensive comparison between our approach and existed mitigation methods. In the future, we intend to explore supplementary performance metrics and extend our analysis to incorporate additional mitigation techniques for a more comprehensive evaluation.

\paragraph{External Validity}
To ensure an equitable comparison with cutting-edge bias mitigation techniques, we leverage a diverse array of real-world models, datasets, and evaluation scenarios. Particularly, we utilize a practical benchmark comprising 16 real-world models thoughtfully curated by prior research~\cite{biswas20machine}. Then, these meticulously chosen models undergo evaluation using four extensively studied datasets in the fairness literature~\cite{bogen2018help, tramer2017fairtest, udeshi2018automated}. We conducted experiments under identical setups and subsequently validated our findings~\cite{biswas20machine}. In addition to assessing \fairautoml against alternative methods within our established settings and benchmarks, we subject \fairautoml to evaluation using the Parfait-ML~\cite{tizpaz2022fairness} benchmark, a leading-edge bias mitigation framework.

\paragraph{Internal Validity} 
Implementing \fairautoml on top of \as may introduce a threat to its actual bias mitigation performance. In other words, the favorable outcomes achieved by \fairautoml could be attributed to its integration with \as. To address this threat, we evaluated \as on various benchmarks, comparing its performance with (\fairautoml) and without (\as) our proposed approaches, to gauge the effectiveness of \fairautoml.

\section{Conclusion}
\label{sec:conclusion}
We present \fairautoml, an innovative system that enhances existing AutoML frameworks to resolve fairness bugs. The core concept of \fairautoml is to optimize the hyperparameters of faulty models to resolve fairness issues. This system offers two novel technical contributions: a dynamic optimization function and a search space pruning approach. The dynamic optimization function dynamically generates an optimization function based on the input, enabling AutoML to simultaneously optimize both fairness and accuracy. The search space pruning approach reduces the size of the search space based on the input, resulting in faster and more efficient bug repair. Our experiments show that \fairautoml outperforms \as and conventional bias mitigation techniques, with a higher rate of bug repair and a better fairness-accuracy trade-off. In the future, we plan to expand the capabilities of \fairautoml to include deep learning problems, beyond the scope of the current study.

\section{Data Availability}
\label{sec:data}
To increase transparency and encourage reproducibility, we have made our artifact publicly available. All the source code and evaluation data with detailed descriptions can be found here~\cite{fazenodo}.


\begin{acks}
This work was supported in part by National Science Foundation under Grant  CCF-15-18897, CNS-15-13263, CNS-21-20448, CCF-19-34884, and CCF-22-23812. Additionally, our sincere gratitude extends to the reviewers for their invaluable insights, which significantly contributed to enhancing the quality of the paper.
\end{acks}

\balance
\bibliographystyle{ACM-Reference-Format}
\bibliography{refs}


\end{document}